# Perovskite PV-powered RFID: enabling low-cost self-powered IoT sensors

Sai Nithin R. Kantareddy, Ian Mathews, Shijing Sun, Mariya Layurova, Janak Thapa, Juan-Pablo Correa-Baena, Rahul Bhattacharyya Tonio Buonassisi, Sanjay E. Sarma and Ian Marius Peters

*Abstract*— Photovoltaic (PV) cells have the potential to serve as on-board power sources for low-power IoT devices. Here, we explore the use of perovskite solar cells to power Radio Frequency (RF) backscatter-based IoT devices with a few µW power demand. Perovskites are suitable for low-cost, high-performance, low-temperature processing, and flexible light energy harvesting that hold the possibility to significantly extend the range and lifetime of current backscatter techniques such as Radio Frequency Identification (RFID). For these reasons, perovskite solar cells are prominent candidates for future low-power wireless applications. We report on realizing a functional perovskite-powered wireless temperature sensor with 4 m communication range. We use a 10.1% efficient perovskite PV module generating an output voltage of 4.3 V with an active area of 1.06 cm$^2$ under 1 sun illumination, with AM 1.5G spectrum, to power a commercial off-the-shelf RFID IC, requiring 10 - 45µW of power. Having an on-board energy harvester provides extra-energy to boost the range of the sensor (5x) in addition to providing energy to carry out high-volume sensor measurements (hundreds of measurements per min). Our evaluation of the prototype suggests that perovskite photovoltaic cells are able to meet the energy needs to enable fully autonomous low-power RF backscatter applications of the future. We conclude with an outlook into a range of applications that we envision to leverage the synergies offered by combining perovskite photovoltaics and RFID.

## I. INTRODUCTION

Photovoltaic (PV) energy harvesters have the potential to enable power-autonomous Internet of Things (IoT) sensors that can operate for years without any need to replace batteries [1]. Recently, perovskite-based photovoltaics have evolved as potential low-cost, high-performance, low-temperature processable, and flexible light energy harvesters [2]. Combining such a low-cost energy harvesting technology with low-power wireless sensor technologies will reduce the constraints in deployment and maintenance (by eliminating battery replacements), power availability and the communication range of sensors in industrial IoT and wireless environmental monitoring applications [3]–[5].

Advances in low-power electronics and communication techniques have enabled a variety of low-power wireless sensing technologies leveraging Radio Frequency (RF) backscattering without any active radio components. Passive Radio Frequency Identification (RFID) is the most prominent RF-backscattering based wireless technology, with globally accepted standards, and is widely used in industrial automation and supply chain applications. RFID (and similar RF-backscatter techniques) consumes ultra-low power (a few µWs) compared to active radio techniques and harvests the incident RF signals to meet this energy demand. However, the purely RF-based power and communication links constrain the device's communication range to a few meters and data-rate to 100s of kbps due to the limited total RF signal strength (1W at the source). For example, longer communication ranges require more power available for backscatter or a slower data rate, similarly, reducing the device's energy consumption increases the energy available for higher data-rate and longer communication range. Passive RFID uses a single communication channel for RF power and data-transmission, thus the communication range suffers (limited to a few meters). New techniques such as LoRa backscatter [6] promise extended ranges (100s of m), using separate channels for power and data-transmission, at slower data-rates, but still in the nascent stages of development. We propose an alternate solution to increase the communication range by using perovskite photovoltaics as on-board energy harvesters that decouple power and communication links—thereby potentially enabling low-cost long-range battery-less IoT devices.

Large scale outdoor PV energy production is generally achieved using well-established Si-based PV technology due to its balance of suitable efficiency and price point. Using PV as a power source for RFID type devices requires additional

Sai Nithin R. Kantareddy is with the Department of Mechanical Engineering, Massachusetts Institute of Technology, MA- 02139. E-mail: nithin@mit.edu
Ian Mathews is with the MIT PV Lab and the Department of Mechanical Engineering, Massachusetts Institute of Technology, MA- 02139. E-mail: imathews@mit.edu
Shijing Sun is with the MIT PV Lab and the Department of Mechanical Engineering, Massachusetts Institute of Technology, MA- 02139.
Janak Thapa is with the MIT PV Lab and the Department of Mechanical Engineering, Massachusetts Institute of Technology, MA- 02139
Mariya Layurova is with the MIT PV Lab and the Department of Mechanical Engineering, Massachusetts Institute of Technology, MA- 02139
Janak Thapa is with the MIT PV Lab and the Department of Mechanical Engineering, Massachusetts Institute of Technology, MA- 02139

Juan-Pablo Correa-Baena is with the School of Material Science and Engineering, Georgia Institute of Technology, GA- 30332
E-mail: jpcorrea@gatech.edu
Tonio Buonassisi is with the MIT PV Lab and the Department of Mechanical Engineering, Massachusetts Institute of Technology, MA- 02139. E-mail: buonassisi@mit.edu
Ian Marius Peters is with the MIT PV Lab and the Department of Mechanical Engineering, Massachusetts Institute of Technology, MA- 02139. E-mail: impeters@mit.edu
Sanjay E. Sarma is with Auto-ID Labs and the Department of Mechanical Engineering, Massachusetts Institute of Technology, MA- 02139. E-mail: sesarma@mit.edu

attributes beyond efficiency such as low-cost integration, light-weight modules, mechanical flexibility, tunable opacity, etc [7]–[9]. Perovskite PV technology offers, at least in principle, these additional benefits, making it a highly interesting candidate for the integration of PV with RFID. However, it is important to note that other emerging thin film PV technologies such as CdTe, organic PV, amorphous-Si, GaAs are also interesting candidates for powering IoT devices [10]–[13]. Perovskites can potentially enable multi-year system lifetimes, however, the current environmental stability (due to temperature and humidity) of the materials is low. Many research groups are currently working on reducing the material degradation rate with new chemical composition and encapsulation techniques [14]. Table 1 summarizes the major synergies of combining perovskite PV with RF-backscatter sensors compared to alternate power sources such as Li battery, and RF, Piezoelectric or thermoelectric [15] energy harvesters. PV is a high energy density source, which reduces the footprint of the energy harvester required to meet IoT device's power demand. It is important to note that thermoelectric energy harvesters are also capable of high energy densities [16], for example, a conventional 40 mm x 40 mm thermoelectric generator can harvest enough energy to power soil moisture sensors. Additionally, rapid advances in efficiencies of perovskite PV cells from 3.81 % to 23.7 % in just 9 years has positioned perovskite materials as promising in developing future high-performance and low-cost photovoltaic cells [17]. Perovskites have the potential for roll-to-roll manufacturing on flexible substrates, which makes integration with RFID tags highly economical and practically feasible as compared with any other PV technology. Scalable perovskite–RFID integration can enable long-range wireless sensing tags without significantly increasing the current passive tag price (7-15 cents).

Research has extensively been conducted to develop perovskite PV cells with strong performance both outdoors and indoors. Few perovskite PV-powered devices such as textile based flexible perovskite PV powered LED [18] and perovskite PV powered electrochromic batteries for smart windows [19] are already shown in the available literature, but there are no reports of fully functional wireless sensor prototypes. Herein, we present a perovskite PV-RFID sensor, with an onboard digital temperature sensor, that can wirelessly transmit temperature data to a reader within a distance of 4 m under sufficient illumination. Figure 1 shows a graphical illustration of our device where the power and the communication links are decoupled: energy converted from light using perovskite PV (layer-wise device architecture is also shown in the figure) powers the Integrated Circuit (IC) and the RF signal is used as the communication link. The reader acts as a gateway between hundreds of such nodes and the internet. Our device operates in the standard unlicensed 902-928 MHz band, meaning it is scalable with the existing infrastructure, which is prominently used for the industrial, scientific and medical applications in the United States. The same energy harvesting mechanism can be still used in other bands, for example, EU and Asian countries, and it is independent of the regulations. In this paper, we show how the perovskite PV cell and RFID sensor are integrated into a single system. We show how the perovskite solar cell meets the energy needs of our device and extends the range by 5 times. We also present an outlook into few potential perovskite PV-RFID applications and the synergies offered by this combination.

This article is organized as follows: In Section II, we discuss the materials, fabrication steps and integration of the energy harvester module with the wireless sensor. In Section III, we discuss the results of testing the functional prototypes as well as the factors impacting the performance of the devices. In

Table 1: Advantages of perovskite PV over other potential power sources for RF backscatter sensors

| Desired features | Potential external energy sources for RF-backscatter sensors | | | | | |
|---|---|---|---|---|---|---|
| | Perovskite PV | Silicon PV | Lithium battery | RF | Piezoelectric | Thermoelectric |
| Thin form factor (< 5 mm) | ✓ | ✗ | ✓ | ✓ | ✓ | ✓ |
| Plastic inlays | ✓ | ✓ | ✗ | ✓ | ✓ | ✓ |
| Integration with roll-to-roll manufacturing | ✓ | ✗ | ✗ | ✓ | ✓ | ✗ |
| Cost competitiveness with 10 cents sensor | ✓ | ✗ | ✗ | ✓ | ✗ | ✗ |
| Energy density | High | ✓ | High | Low | Low | High |
| Full/Semi-transparent | ✓ | ✗ | ✓ | ✗ | ✓ | ✓ |
| Ubiquitous environment | ✓ | ✓ | ✓ | ✓ | ✗ | ✗ |
| Lifetime of the system | Potentially few years | > 10 Yrs | 1-2 Yrs | > 10 Yrs | > 10 Yrs | > 10 Yrs |
| References | [7], [8], [31] | [32] | [33] | [34] | [35] | [36][15] |

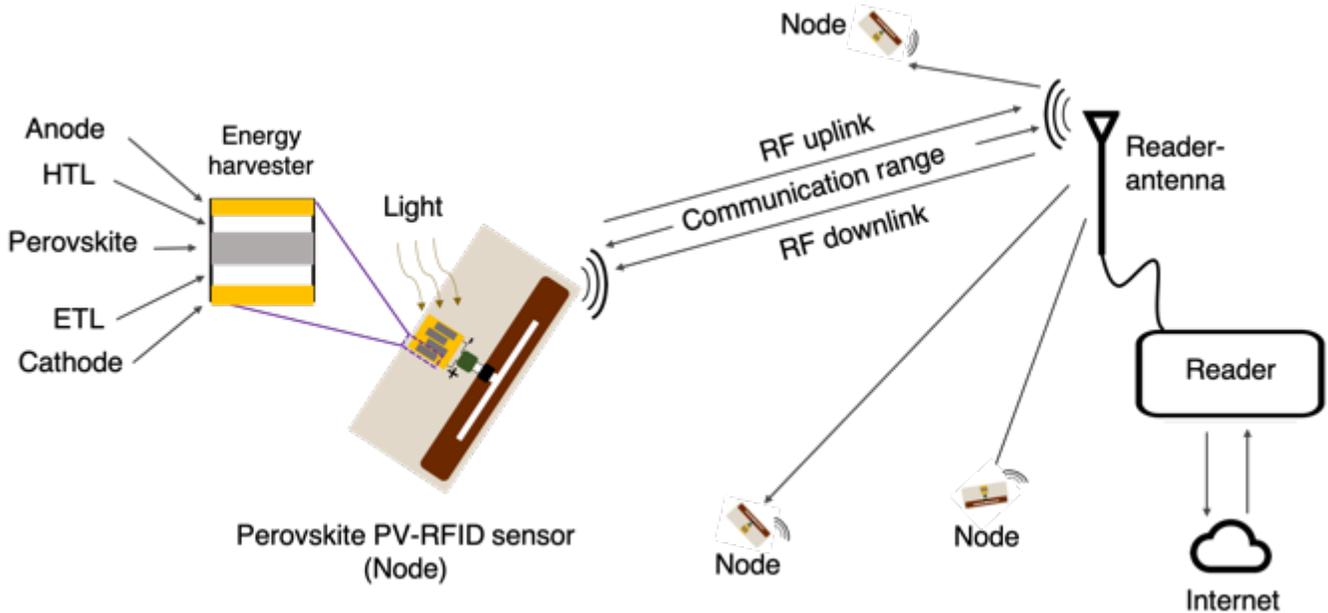

Figure 1: Illustration of perovskite PV-RFID wireless network; Layout of the device showing the material architecture of perovskite energy harvester. ETL and HTL refer to electron transport layer and hole transport layer, respectively

Section IV, we present few potential applications for these sensors and how perovskite PV-RFID offers unique advantages as a single platform.

## II. Energy Harvester

### A. Perovskite solar cell device architecture and fabrication

Following our previously reported methods, we fabricated perovskite solar cells using the state-of-the-art deposition recipe for $Rb_{0.01}Cs_{0.05}(FA_{0.83}MA_{0.17})_{0.94}Pb(I_{0.83}Br_{0.17})_3$ with excess $PbI_2$ added for passivation following our previously reported methods [27]. The layer-by-layer device architecture is shown in Figure 2 (a) along with the fabricated device. Detailed fabrication steps and the roles of each layer are outlined in the additional information section. The optical band gap of our perovskite material is 1.6 eV, which is larger than the traditional $CH_3NH_3PbI_3$ band gap (1.5 eV)[21], leading to higher voltages desired for powering low-power IoT eleactronics.

Figure 2 (b) shows the current-voltage (IV) relationship for our fabricated device under one sun illumination with a photoactive area of 1.06 cm$^2$ and an overall area of 2.8 cm$^2$. We measured the device's power conversion efficiency considering the photoactive area as 10.1 % at 3.7 mA cm$^{-2}$ short-circuit current density ($J_{sc}$), 4.3 V open-circuit voltage ($V_{oc}$), and 0.6 Fill Factor (FF). In addition to Voc and Isc, FF is another parameter, defined as the ratio of the maximum power from the solar cell to the product of Voc and Isc, that determines the maximum power output. Graphically, FF is the measure of the "squareness" of the IV curve (solid line in Figure 2 (b)). Further optimizing the fabrication process and perovskite recipe can yield higher efficiencies. For our purpose, a 10.1 % efficient cell is sufficient to solve the on-board energy demand of the sensors, therefore, fabricating high efficiency modules is not our objective in this study. However, the device's efficiency calculated by considering the total exposure area is 3.5 %, which requires improvement as the reduction in device footprint will reduce the manufacturing costs. Moreover, improving voltage is more important than improving current, because the current, even in the sub-optimal device used here, exceeds all needs. Our target voltage for the device is 3 V at the peak power point for charging the capacitor, which is achieved by connecting 4 of the perovskite solar cells in series ($V_{oc}$'s of individual cells add up)

We evaluate the performance of the cell under different wavelengths of light by measuring the External Quantum Efficiency (EQE) as shown in Figure 2 (c). Measurements show high values (> 70%) in the 400-750 nm range, which is consistent with light available from a variety of sources (sun, LED and other lighting). An interesting feature of perovskite materials is the band gap tunability to optimize the light harvesting efficiency for specific environmental conditions (for specific lighting conditions). For example, perovskite cells' bandgap can be tuned to derive high performance in low-light indoor conditions (under LED, fluorescent lighting, etc. [22]).

### B. Integration with RFID

For the RFID tag, we made use of the commercially available EM 4325 IC containing an on-board digital temperature sensor. The device's schematic is shown in Figure 2 (d) and consists of an antenna T-matched [23] to the IC's impedance in 902-928 MHz frequency, and the perovskite PV module connected to power the IC. The device's schematic is shown in Figure 2 (d). All components are assembled on a plastic substrate. However, it is important to note that the perovskite cells are currently fabricated on glass and fabricating perovskites on plastic is still an ongoing research topic in this field. There are also startups like Saule Technologies exploring the commercialization of printed perovskite cells. In this work, the prototype is realized with perovskite on glass with tag and the capacitor on plastic. The tag communicates its ID and other information from ICs memory bank (when the reader issues a read command) by encoding information (0 s and 1s) on to the modulated backscattered signal. The backscattered signal is generated by merely reflecting the incident RF radiation from the reader system, therefore, the tag does not require any active transreceiver. Information is encoded on to this signal by switching IC's its internal impedance following the rules dictated by globally standard EPC Gen 2 protocol. Our device's performance is evaluated by acquiring the sensor measurements using standard industrial RFID equipment. For the purposes of these tests, we made use of an Impinj Speedway RFID reader connected to a circularly polarized antenna with a gain of 8.5 dBi and separated from the tag by a distance of 1 m.

write, sensor measurements, etc. Average current consumption data, taken directly from the manufacturer, for different operation modes of the IC, are presented in Figure 3 (a). IC can be put in sleep mode, consuming the lowest current (1.6 µA), to conserve on-board energy. In the ready mode, when the IC is ready to receive temperature measurement acquisition commands, the IC consumes an average of 6 µA. However, every temperature measurement requires 5x higher current (30 µA) for 8ms duration. Based on these values, we simulated the overall power consumption of the sensor depending on how heavily the sensor is used (number of sensor measurements per hour). Figure 3 (b) shows this in a plot, for example, 20,000

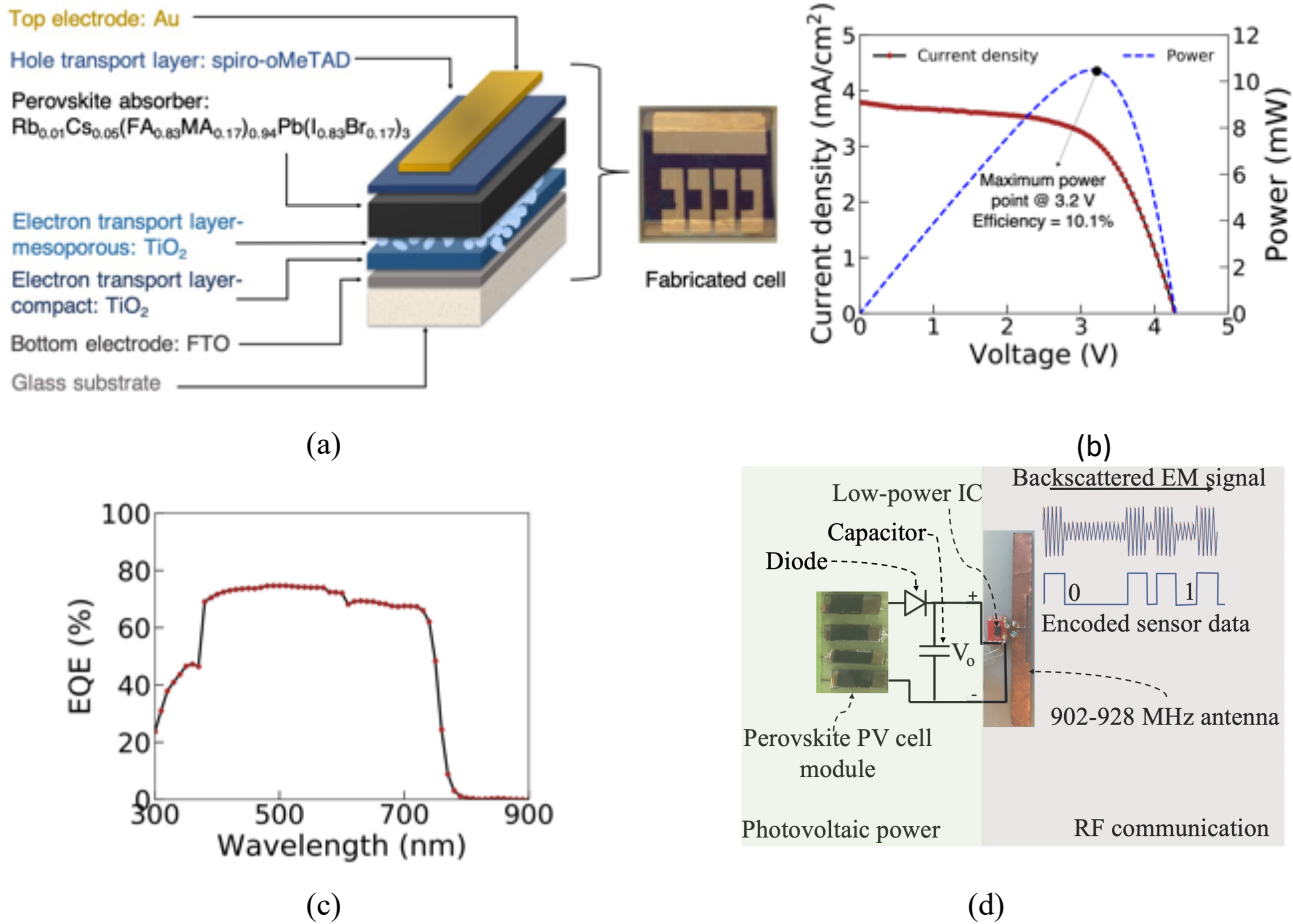

Figure 2: Perovskite PV cell architecture (a), current density-voltage curve of the device (b), plot of EQE of the cell (c), perovskite PV-RFID sensor integration showing the PV power and RF communication paths (d)

### III. RESULTS AND DISCUSSION

RF-backscatter tag ICs require extremely low-power (few µW) to bootup and backscatter the modulated signal with encoded information. Currently, available ICs are designed to harvest an equivalent portion of the incident RF signal to power themselves (IC sensitivity ranging from -8.3 dBm to -22 dBm). However, by providing external energy through perovskite PV, maximum RF signal (without any consumption at the IC) is backscattered--thereby increasing the communication range. ICs typically can operate in several modes such as sleep, ready, measurements per hour require around 20 µW power (computed by taking constant 6 µA of current in ready state and 30 µA for 8ms during each temperature measurement at 1.5V input voltage). We compared this power demand with the power harvesting potential for perovskites under outdoor and indoor conditions. For the record perovskite efficiencies reported in the literature [24], [25], 2 mm$^2$ large perovskite can harvest 30 µW in outdoor conditions in Boston, MA and a 200 mm$^2$ large perovskite module can harvest 20 µW under indoor fluorescent lighting. These results show that small area modules can sufficiently meet the power demand by PV-RFID sensors

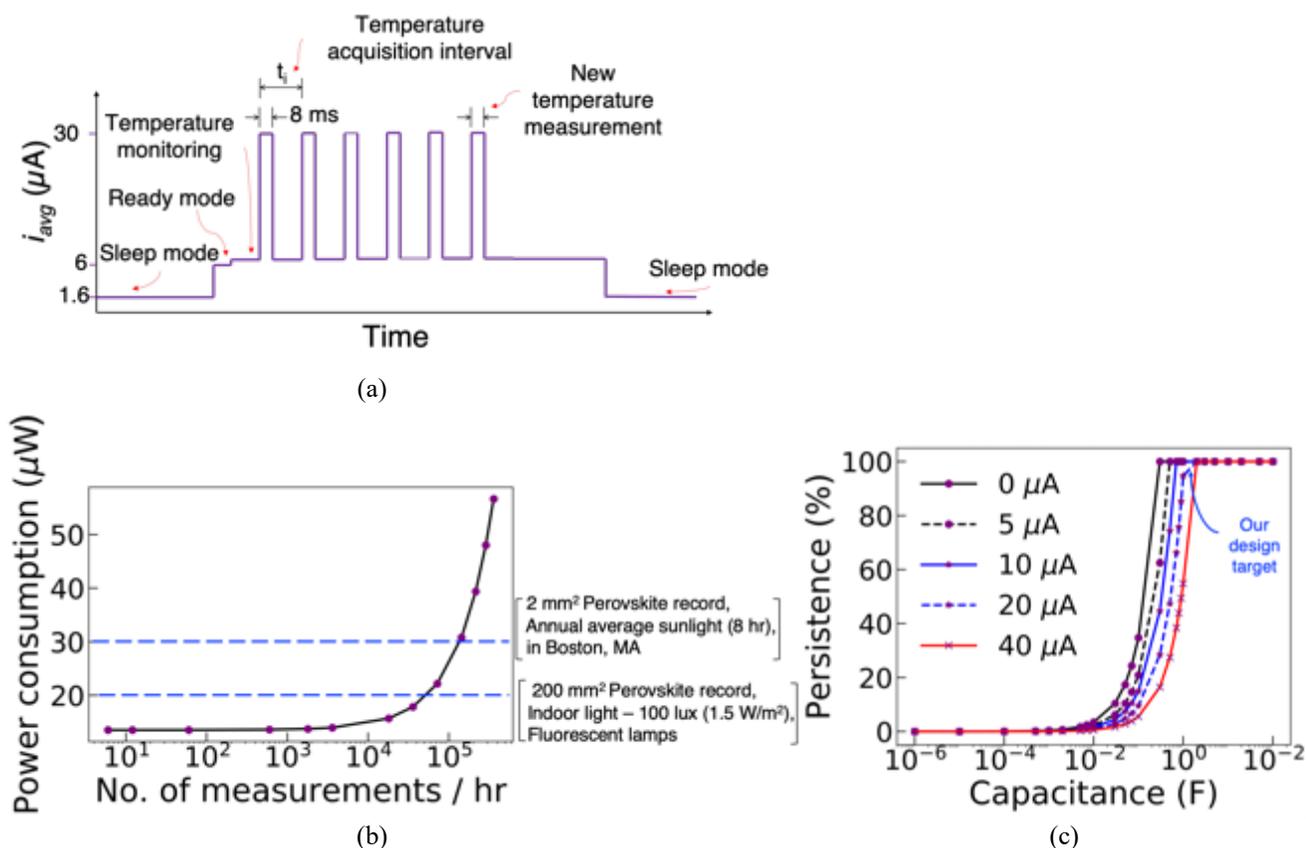

Figure 3: Average current consumption for different operation modes (a), average current consumption for different interrogation rates (b), and device's persistence at different capacitances and current leaks (c)

at high temporal data resolution modes. We use an external capacitor as an energy buffer to store the energy on-board during low-light conditions as a backup power supply to the device. However, the charge leakage in capacitors significantly affects the life time of the device. Figure 3 (c) shows simulated results of estimated persistence (availability of the device in a day) of the device at varying leakage currents from 0 to 40 µA for a range of capacitor sizes (1 µF to 100 F) at the maximum charging voltage of 3 V (safe voltage for the IC). Availability of the device is the fraction of the total duration the sensor is available to transmit data. This duration is calculated by comparing the total energy required to take measurements in a day with the total amount of energy harvestable by the cells minus the energy dissipated due to charge leak in the capacitors. To minimize costs and footprint of the device, we selected 1 F capacitor (commercially available, AVX corporation) that is estimated to provide 100 % persistence when charged fully to 3 V. Also, brand new capacitors tend have higher leakage currents than aged capacitors, therefore, the obtained results could be improved by appropriately aging the capacitors. However, for applications where sensor measurements are made only when the light is available (e.g. moisture sensors for precision agriculture), an external buffer capacitor is not necessary and perovskite PV can be directly connected to the IC, which further reduces the overall cost of the device. Additionally, developing system-on-chip solutions for perovskite PV with on board tiny energy buffers will also reduce the need for external capacitors.

From the test results, using a solar simulator with one sun intensity (AM 1.5G spectrum), the device shows sharp charging times and takes around 300 seconds to reach the threshold voltage (1.5 V for read and EEPROM operations on the IC) required to boot up the IC. The capacitor fully charges to a 3 V limit in 50 mins (see Figure 4 (a), while simultaneously powering the device (see temperature measurements in Figure 4 (b)). The device can function in the semi-passive mode while simultaneously being charged, therefore, the interrogator need not wait till the capacitor is fully charged. Once the equilibrium is reached (the terminal voltage negligibly increases thereafter), the light source is switched off to measure the device's discharge rate. As shown in Figure 4 (a), the device discharges at a slower rate than its charging rate. Terminal voltage drops from full 3 V to 2 V in 5000 seconds and discharges to threshold limit (1.5 V) in 8000 seconds. Therefore, around 45 min light availability powers the wireless sensor prototype for a total of 185 min (4 x time). Capacitor's self-discharge currents significantly affect the persistence level, therefore, a higher capacitance with low self-discharge currents is an ideal design choice. Figure 4 (b) shows the plot of temperature measurements obtained while the device went through the charging and discharging cycles in Figure 4 (a). This shows that the prototype is simultaneously taking and transmitting temperature measurements while the capacitor is being charged.

We compared the read range of our perovskite PV-RFID device with a purely passive RFID device using Voyantic's

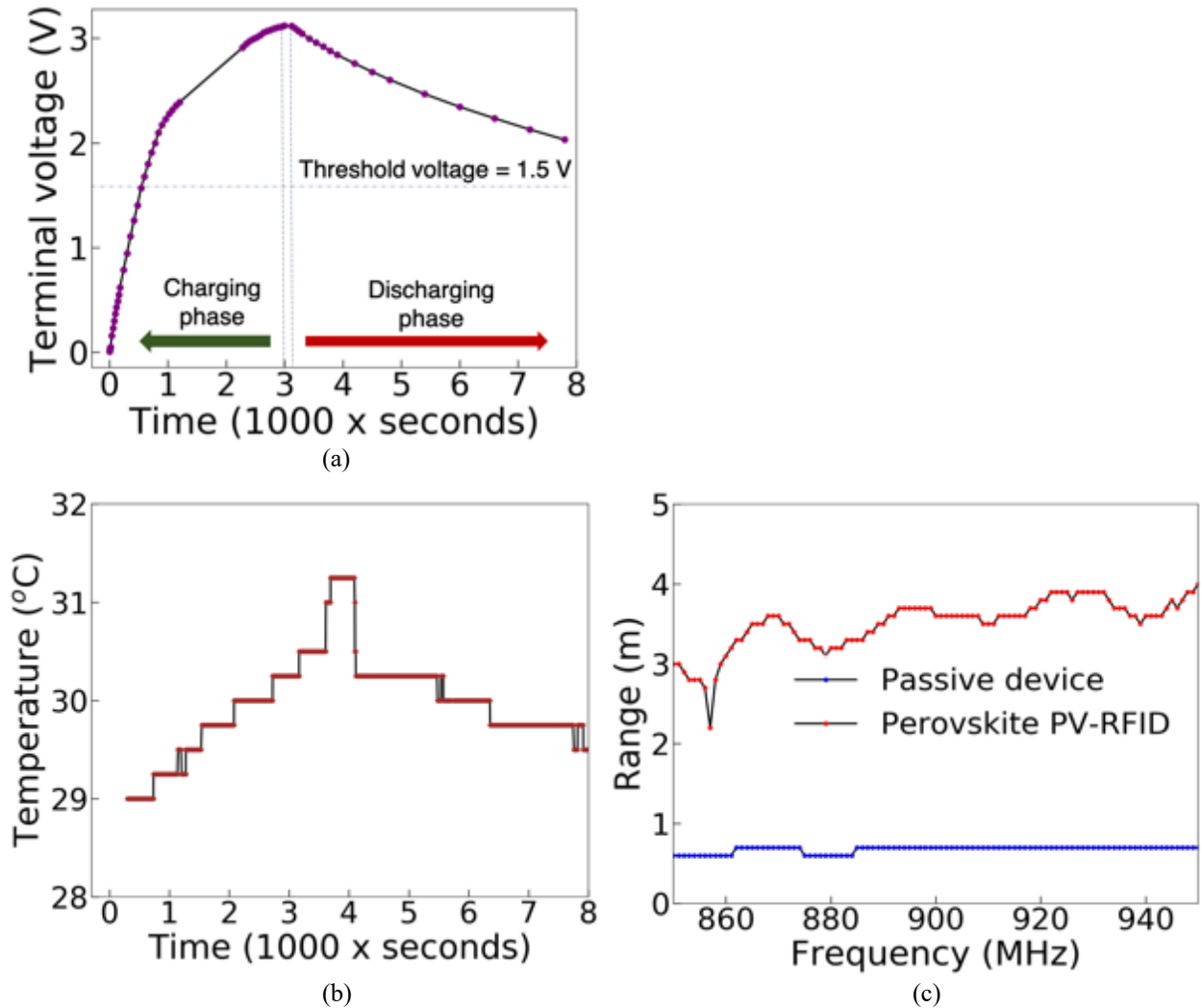

Figure 4: Charge and discharge cycles in the device (a), temperature data from the sensor during the charging and discharging cycles (b), and range improvement in perovskite PV-RFID compared to passive RFID sensor

Tagformance Pro setup. The tag is placed at a known distance (equivalent to one wavelength) from the Tagformance's antenna, threshold RF power (minimum RF power required to boot up the IC) is measured as a function of frequency over a wide band from 850 to 950 MHz. Tagformance then calculates the read range indirectly from RF power measurements and tag-antenna separation. As shown in Figure 4 (c), a purely passive device can transmit over a range of less than 1 m, whereas the perovskite PV-RFID device of the same antenna can transmit to around 4 m distance. Hence, the perovskite PV-RFID device shows a 5x boost in range with 0.9 m variance. This device is a generic prototype, without optimizing the antenna for a specific application, however, communication range can be further increased by optimizing the antenna's design (for antenna's gain and transmission co-efficient parameters) using high-frequency electromagnetic simulations considering the environment and surrounding dielectric materials in play. Theoretically, 10s of m range can be achieved as estimated in [10]. The trend in increasing read range with energy harvesters is also comparable to the boost in range found in available literature [26]. In addition, Perovskite photovoltaics provide additional benefit of realizing this potential at a fraction of the cost with advantages of mechanical flexibility, optical transparency, etc. Long-range sensors reduce the need for extensive physical gateways/readers and antennas in the field—thereby reducing the overall infrastructure cost needed to set up.

IV. APPLICATION ANALYSIS

Perovskite PV powered RF backscatter sensors (RFID or any other emerging protocols) will have numerous potential applications, for example, as sensors embedded in car windshields [27], long-range soil moisture sensors for precision agriculture, asset trackers in supply chain, and battery-less indoor sensors [28]–[30]. These applications will benefit from the perovskite – RFID integration's unique advantages such as transparent and flexible features, ubiquitous energy harvesting in different environmental conditions (indoors and outdoor lighting conditions), long-communication ranges, low-cost manufacturability and scalable battery-less sensing. Figure 5 shows the landscape of few potential applications requiring diverse set of these desired features. For example, windshield sensors need transparent, outdoor performance, long range and sparse scale features, on the other hand, sensors for supply chain need low-light energy harvesting, low-cost, flexible, highly-scalable and moderate range features. Energy harvesting requirements and sizing of the PV cell and capacitor would change depending on the application. For example, soil moisture sensors will be used outdoors with reader interrogating once or twice a day, therefore, requires smaller PV cell area and smaller energy storage. On the other hand, indoor battery-less sensors require larger PV cell area with higher band gap to harvest low intensity light and larger energy storage capacity to provide energy backup throughout the night. Depending on the use case, the components on our device can be customized accordingly. Thus, perovskite PV-RFID sensors could function as one sensor platform usable in multiple different application scenarios.

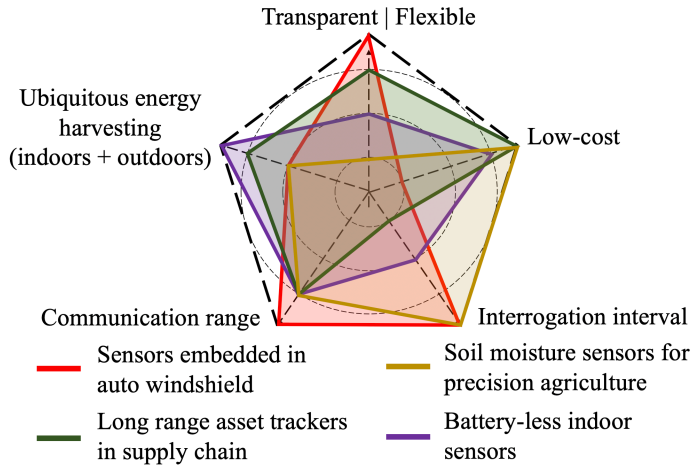

Figure 5: Landscape of few potential applications w.r.t Perovskite PV-RFID related performance metrics

Although perovskites offer numerous excellent synergies to combine with low-cost backscatter sensors such as RFID, environmental stability (due to temperature and humidity) and environmental hazard (due to lead content in the material) are two major problems that require further research and development. Many research groups are currently working towards developing lead-free perovskite cells [30] and encapsulation techniques to prevent material degradation [14]. Lead-free perovskites currently have lower efficiency than lead-containing ones, so in our paper we worked with the state-of-the-art recipe. Any commercialization effort in real-world would aim to use lead-free perovskites. Another area to investigate is the RF dielectric properties of perovskite PV materials in the high frequency ranges (915 MHz, 2.4 GHz, 5 GHz and beyond). High-frequency material characterization will enable us to optimize antenna's radiation capabilities with respect to background dielectrics (induced by perovskite and substrate materials) to derive maximum communication ranges depending on the operating frequency. Progress in these areas can potentially address the implementation concerns on these materials to enable low-cost long-range perovskite PV-powered sensors that can be deployed for years without requiring any human intervention

## V. Conclusion

Low-cost energy harvesters will enable the design of power-autonomous wireless IoT sensors that persist for years without any need for human access or maintenance. We introduce perovskite PV-RFID sensors that leverage the low-cost, high performance and flexible mechanical properties of perovskites and the low-cost, scalable, low-power nature of RF backscattering to enable inexpensive, battery-less IoT devices.

We reduce the idea to practice by developing a functioning prototype of a perovskite-powered wireless temperature sensor. We use a 10.1 % efficient, 1.06 cm$^2$ active area perovskite PV module to power our commercial off-the-shelf RFID IC with on-board digital temperature sensor. The device is capable of high temporal resolution measurements at few (up to 4) meters distance, a 5-fold enhancement compared to traditional passive RFID sensors, due to by decoupling power and communication links. The device features steep charging cycles and slow-discharging cycles, which are favorable patterns for battery-less sensors in environments with intermittent light availability. The extremely low-power demand (10μW) of RFID only requires small area perovskite cells—thereby reducing the overall footprint and material required.

Integrating low-cost RFID manufacturing with roll-to-roll manufacturing of perovskite PV modules has the potential to make perovskite PV-RFID sensors highly cost competitive, compared to battery-based IoT sensors with applications requiring few meters of range. These devices could find use in varying application ranges such as auto windshield sensing, soil moisture sensing, asset tracking, and battery-less indoor sensors that can benefit from the mechanical flexibility, material's band gap tunability, low-cost manufacturing, long-range communication and scalable protocol of PV-RFID technology.

## VI. Acknowledgement

Authors would like to acknowledge the sources of funding for this work. S.N.R.K. has received funding from GS1 organization through the GS1-MIT AutoID labs collaboration. I.M. has received funding from the European Union's Horizon 2020 research and innovation programme under the Marie Skłodowska-Curie grant agreement No. 746516. I.M.P. was financially supported by the DOE-NSF ERF for Quantum Energy and Sustainable Solar Technologies (QESST) and by funding from Singapore's National Research Foundation through the Singapore MIT Alliance for Research and Technology's "Low energy electronic systems (LEES)" IRG.

Electronic supplementary Information

*A. Device Fabrication*

Etched FTO glass (10 Ω/sq) was cleaned in Hellmanex (2% in water) followed by deionized water and ethanol for 15 min at each step. Further cleaning of UV ozone treatment for 15 min was then employed. For the electron-transport layer, a compact $TiO_2$ layer was grown by depositing solution of the titanium diisopropoxide bis(acetylacetonate) 75% wt. (Sigma-Aldrich) in ethanol (1:10 V/V ratio) using spray pyrolysis method at 500°C. A $TiO_2$ mesoporous layer was spincoated on top of the compact $TiO_2$ layer by mixing 1:5 W/W of $TiO_2$ paste (SureChem, SC-HT040): solvent mix (3.5:1 W/W of terpineol: 2-methoxy ethanol). In mesoporous configuration, scaffold structures of the mesoporous materials such as TiO2 help in crystallization of the Perovskite material. These structures define the grain size and crystal orientation during the crystallization process. The substrate was then heated at 500°C for 1 hour to remove the organic solvent. The perovskite precursor solution was prepared by mixing FAI (1 M, Dyenamo), $PbI_2$ (1.1 M, TCI), MABr (0.2 M, Dyenamo) and $PbBr_2$ (0.22 M, TCI) in a 9:1 (v:v) mixture of anhydrous DMF:DMSO (Sigma Aldrich). We then added solutions of CsI (Sigma Aldrich) and RbI (Signma Aldric) respectively, both prepared as 1.5 M stock solution in DMSO, in a 5:1:95 volume ratio of CsI:RbI:perovskite solution. The above precursor solution was spincoated onto the substrate using a two-step spin coating program (10 s at 1000 rpm, 20 s at 6000 rpm). 150 μL of chlorobenzene was added as anti-solvent during the second step. The films were then annealed at 100 °C for 20 min. Spiro-OMeTAD (2,2′7,7′-tetrakis-(N,N-di-p-methoxyphenyl amine)-9,9′-spirobifluorene, LumTec LT-S922) was used as the hole-transport layer. For every gram of spiro-OMeTAD, 227 μL of Li-TFSI (Sigma-Aldrich, 1.8 M in acetonitrile) solution, 394 μL of 4-tert-butylpyridine (Sigma-Aldrich) solution, 98 μL cobalt complex (FK209, Lumtec, 0.25 M tris(2-(1H-pyrazol-1-yl)-4-tertbutylpyridine) cobalt(III) tris(bis(trifluoromethylsulfonyl)imide) in acetonitrile) solution, and 10,938 μL of chlorobenzene was added as dopants. 65 μL of the mixed spiro solution w spincoated onto the perovskite films at 3000 rpm for 30 s. A 100 nm gold top electrode was then deposited on the perovskite cell via thermal evaporation.